\documentclass[twoside]{article}

\usepackage{amsmath,amsfonts}

\newcommand{\ket}[1]{\left|#1\right>}

\newcommand{\ul}{\underline} 
\newcommand{\f}[1]{\mbox{\boldmath$#1$}}

\newcommand{\bea}{\begin{eqnarray}}
\newcommand{\ea}{\end{eqnarray}}
\newcommand{\eea}{\end{eqnarray}}
\newcommand{\ord}{{\cal O}}

\def\beq{\begin{equation}}
\def\eeq{\end{equation}}
\def\bea{\begin{eqnarray}}
\def\eea {\end{eqnarray}}

\begin{document}





\vspace*{0.88truein}



\centerline{\bf HIDDEN SYMMETRY DETECTION ON A QUANTUM COMPUTER}

\vspace*{0.37truein}

\centerline{\footnotesize R.~SCH\"UTZHOLD\footnote{email: 
{\tt schuetz@theory.phy.tu-dresden.de}}}
\vspace*{0.015truein}
\centerline{\footnotesize\it 
Institut f\"ur Theoretische Physik, Technische Universit\"at Dresden}
\baselineskip=10pt
\centerline{\footnotesize\it 
01062 Dresden, Germany}
\vspace*{10pt}
\centerline{\footnotesize W.~G.~UNRUH\footnote{email: 
{\tt unruh@physics.ubc.ca}}}
\vspace*{0.015truein}
\centerline{\footnotesize\it 
Department of Physics and Astronomy, University of British Columbia}
\baselineskip=10pt
\centerline{\footnotesize\it 
Vancouver, British Columbia, Canada V6T 1Z1}
\baselineskip=10pt
\centerline{\footnotesize\it 
Canadian Institute for Advanced Research Cosmology and Gravity Program}
\vspace*{0.225truein}

\vspace*{0.21truein}

\abstract{The fastest quantum algorithms 
(for the solution of classical computational tasks) 
known so far are basically variations of the hidden subgroup problem with 
{$f(U[x])=f(x)$}.
Following a discussion regarding which tasks might be solved efficiently by 
quantum computers, it will be demonstrated by means of a simple example, 
that the detection of more general hidden (two-point) symmetries 
{$V\{f(x),f(U[x])\}=0$} by a quantum algorithm can also admit an exponential 
speed-up.
E.g., one member of this class of symmetries {$V\{f(x),f(U[x])\}=0$} 
is discrete self-similarity (or discrete scale invariance).
\\
PACS: 
03.67.Lx, 
89.70.+c. 
}

\vspace*{10pt}



\vspace*{1pt}
\section{Introduction}\label{Introduction} 

Shor's striking discovery \cite{shor}, that quantum computers could accomplish 
tasks such as factoring large numbers exponentially faster than the best 
(known) classical methods, motivates the quest for further quantum algorithms 
exhibiting an exponential speed-up, see, e.g., \cite{nielsen} for a review.
Together with a number of black-box problems 
\cite{deutsch,jozsa,bernstein,simon}, some of which also admit an exponential 
speed-up, Shor's algorithm can be generalised to the so-called 
``hidden subgroup problem'': given a function $f$ with the property
\bea
\label{subgroup}
\forall\,x,y\;:\;f(x)=f(y)\;\leftrightarrow\;y\in U^{\mathbb Z}x
\,,
\ea
for some transformation $U$, find $U$.
I.e., $f$ is constant on the co-sets 
of the subgroup $\{x,Ux,U^2x,U^3x,\dots\}$ generated by $U$ 
and assumes a different value at each co-set.
(Here we restrict our consideration to the case of one generator $U$ only,
for more than one generators, the situation is analogous.)
For example, in the case of Shor's algorithm, the transformation $U$ is given 
by $U[x]=x+p$, and for Simon's \cite{simon} problem, it is $U[x]=x \oplus p$ 
with $\oplus$ denoting the bit-wise addition modulo two, e.g., 
$1001 \oplus 0101 = 1100$. (Note that $x \oplus p \oplus p = x$.)

Hence, in comparison with classical methods, the number of known quantum 
algorithms which are (as far as we know) significantly faster is tiny -- 
but one might hope that there are many more to be discovered.
The question we wish to examine is: which other problems -- and perhaps further
expansions of the known tasks -- could (also) admit an exponential speed-up?
More precisely, we shall investigate whether there are general features of 
problems which are important for an exponentially fast quantum algorithm,
and give a specific example (in which such a speed-up is accomplished) via
an extension of Simons's and Shor's problem.

In particular we shall consider problems which can be cast into the 
following form: given a function $f\,:\,x\,\to\,f(x)$ on an exponential 
number of arguments $x$, where $f$ is known to possess some property 
(from a given class of properties), find that property -- where the term 
``property'' can refer to any extracted information in general.
Evaluating $f(x)$ on a given arbitrary argument $x$ is assumed to be 
polynomially (in the length of $x$) implementable\footnote{
I.e., the problem to be solved must be at least in {\em PSPACE} -- 
remember that {\em P}\,$\subseteq$\,{\em NP}\,$\subseteq$\,{\em PSPACE},
see, e.g., \cite{nielsen}.}. 
We shall investigate some features \cite{intuitive} of the class of 
properties with an (apparently) exponential speed-up by a quantum 
computer over a classical one.
We shall also show how such an exponential speed-up can be achieved for a 
property we call a hidden symmetry.

Note that our discussion will not be concerned with the use of quantum 
computation to simulate physical systems, nor with the application of 
quantum phenomena to transmit information 
(quantum cryptography or super-dense coding, etc.)
or to extract information from an external physical system
(such as quantum imaging, see, e.g., \cite{imaging}, 
or Elitzur-Vaidman-type problems \cite{bomb}),
i.e., we only consider quantum information processing.

\vspace*{1pt}
\section{Relevance of arguments}\label{Relevance of arguments} 

One aspect which seems \cite{intuitive} to be important for an exponential 
speed-up is the relevance of the arguments $x$ with respect to the property
under consideration.

Typically, sequential \cite{ground} quantum algorithms for solving
problems as described above can be formulated as black-box algorithms 
which can be cast into the following most general form 
\bea
\label{expon}
\ket{\Psi}=
{\cal U}_m\,{\cal U}_f\,{\cal U}_{m-1}\dots
{\cal U}_1\,{\cal U}_f\,{\cal U}_{0}\ket{0}
\,,
\ea
where the unitary gate ${\cal U}_f$ calculates the (black-box) 
function $f$, i.e., 
${\cal U}_f\ket{x}\ket{y}=\ket{x}\ket{y \oplus f(x)}$, 
with the possible extension ${\cal U}_f\to{\cal U}_f\otimes\f{1}$;
and additional unitary operations ${\cal U}_0\dots{\cal U}_m$.
Even if the algorithm originally contained an intermediate measurement, 
it could still be rewritten in this form by using ancilla qubits and 
quantum-controlled operations.

In order to achieve an exponential speed-up the number $m$ as well as the 
realisations of the unitary operations ${\cal U}_0\dots{\cal U}_m$ have to be
polynomial.
Consequently, if the number of arguments $x$ of the function $f(x)$ that 
contain relevant information (for the solution of the problem) is 
exponentially small, then the part of the output state $\ket{\Psi}$ 
corresponding to this relevant information is apparently \cite{intuitive} also
exponentially small, and therefore impossible to extract with a polynomial 
number of measurements.
Supportive (though not conclusive \cite{intuitive}) to this point is 
inserting the identity 
\bea
{\cal U}_f=
\left({\cal P}_{\rm rel}+{\cal P}_{\rm irr}\right)
{\cal U}_f
\left({\cal P}_{\rm rel}+{\cal P}_{\rm irr}\right)
\,,
\ea
into Eq.~(\ref{expon}), where ${\cal P}_{\rm rel}$ and ${\cal P}_{\rm irr}$ 
denote the (orthogonal) projections onto the subspaces of relevant and 
irrelevant arguments $x$, respectively.
Assuming that the unitary operations ${\cal U}_0\dots{\cal U}_m$ do not favour 
\cite{intuitive} the subspace spanned by ${\cal P}_{\rm rel}$ (we do not know 
in advance which arguments $x$ are going to be important and which not) 
the norm of (the sum of) all terms containing at least one ${\cal P}_{\rm rel}$
is exponentially small for the unitary operations are norm-preserving.

As a result, a function with an exponentially small number of relevant 
arguments $x$ does not seem suitable for an exponential speed-up.
Of course, this feature crucially depends on the particular way of encoding the
problem to be solved by a function -- e.g., a function defined as $f(x)=1$ if 
$x$ is a factor of $y$ and $f(x)=0$ otherwise would not be the best choice for
factoring \cite{bennett}.
One should also bear in mind that the above arguments do not exclude
polynomial speed-up -- the Grover search routine \cite{grover}
achieves a quadratic speed-up by exploiting the bilinear structure of
quantum theory, i.e., the normalisation by $1/\sqrt{N}$ instead of $1/N$.

The task of period-finding, for example, where all arguments $x$ are
equally relevant for the solution, is therefore indeed 
(as it should be) a good candidate for an exponential speed-up by a
quantum algorithm. 
As counter-examples 
(which are probably not good candidates for an exponential speed-up), 
we may quote the usual form of the travelling salesman problem 
[with $x$ being one particular route and $f(x)$ the associated length] 
or the task of evaluating the position in chess\footnote{
If $x$ denotes one possible continuation of the game (a so-called ``line'')
and $f(x)$ the outcome (win, loss, or draw) then the vast majority of 
arguments $x$ are irrelevant for accessing the position because almost all 
lines with random moves are completely uninteresting.} 
where {\em a posteriori} almost all arguments $x$ are completely irrelevant 
-- but we do not know {\em a priori} which.

\vspace*{1pt}
\section{Excess information}\label{Excess information}

Since every quantum computation is (at least in principle) unitary and hence 
reversible, it is impossible to lose any information during this process -- 
except by the (final) measurement (e.g., the phases are lost) or by 
transferring the information from the quantum system (computer) alone to its 
entanglement with the ``environment''.
We want to extract only a certain property of the function $f$ -- 
other details of $f$ are irrelevant -- and, therefore, we have to find a way 
to dispose of this excess information.
For example, in Shor's problem $f(x)=f(x+p)$, we only want to know the period
$p$, and not any other details of $f$.
After the measurement of the register $\ket{f}$ one is left with the state 
\bea
\label{period}
\ket{\Psi}=\frac{1}{\sqrt{L}}\sum\limits_{l=0}^{L-1}\ket{x_0+lp}
\,.
\ea
This state contains very little information -- basically just the starting 
point $x_0$ and the period $p$ -- which, after a quantum Fourier transform, 
determine the phase and the value of the wave-number, respectively. 

Of course, here we have to explain the phrase ``very little information''. 
To this end we introduce the notion of the ``classical information of
a quantum state $\ket{\Psi}$'' as the information required to
reproduce the state $\ket{\Psi}$ starting from the state $\ket{0}$ in 
the computational basis via elementary operations \cite{intuitive}.
Note that this notion is obviously not a unitarily invariant (quantum) 
information measure (as $\ket{\Psi}$ is still a pure state).
But since we want to speed up the solution of classical problems, we
should consider the involved quantum operations from a classical point
of view.

In summary, we arrive at an (admittedly rather vague \cite{intuitive}) 
additional condition -- ``not too much excess information'' -- for a 
(classical) problem which is supposed to admit an exponential (quantum) 
speed-up.
As a counter-example, we might consider the average invertability check 
(collision problem) of a function $f(x)$ -- i.e., for a given 
(representative) $y$ in the co-domain, how many $x$ satisfy (in average) 
$f(x)=y$. (This problem is relevant for cryptography.)
Although obviously almost all arguments $x$ are equally important, the 
state after the measurement of $f$ apparently still contains too much 
excess information\footnote{Note that our notion ``classical
  information of a quantum state $\ket{\Psi}$'' is different from the
  generalisation of the Kolmogorov complexity to the quantum case
  introduced in \cite{Kolmogorov}. The latter quantity is bounded from
  above \cite{Kolmogorov} and its upper bound of approximately $2n$
  (where $n$ is the number of qubits) would just correspond to the
  information contained in $x_0$ and $p$ for Shors algorithm. In
  contrast, the ``classical information of a quantum state
  $\ket{\Psi}$'' introduced here can exceed this bound by far: For the
  collision problem, the different and independent coordinates $x$
  satisfying $f(x)=y$ typically contain much more information. Hence
  the generalisation of the Kolmogorov complexity proposed in
  \cite{Kolmogorov} cannot be used to discriminate between the two
  cases (Shors algorithm and the collision problem).}  
(the different and independent coordinates $x$ satisfying $f(x)=y$) 
to get rid of \cite{collision}.

\vspace*{1pt}
\section{Hidden symmetries}\label{Hidden symmetries}

As one would reasonably expect, the hidden subgroup problem satisfies the 
above requirements -- all arguments $x$ are equally relevant and the state 
after the measurement of $f$ is basically determined by one starting point 
$x_0$ and the generator $U$ (e.g., $p$) of the subgroup.
This feature is ensured by the existence of the symmetry (\ref{subgroup})
connecting the values of the function $f$ at each two points $x$ and $U[x]$ 
with a certain relation, i.e., $f(x)=f(U[x])$.
In view of the above remarks, one might expect a similar effect for a more 
general hidden two-point symmetry of the form
\bea
\label{hidden}
V\left\{f(x),f(U[x])\right\}=0
\,,
\ea
where $V$ is some relation generalising the equality in the hidden subgroup 
problem (\ref{subgroup}).

Of course, it remains to be shown whether it is possible to design a quantum 
algorithm which determines $U$ and $V$ exponentially faster than classical 
methods.
One of the major benefits of quantum computation is the superposition 
principle allowing us to test all possible values of $x$ at once 
(``quantum parallelism'').
In view of this observation one would expect that it is advantageous to 
represent the symmetry operations in a (somehow \cite{intuitive}) linear 
fashion.
(This seems to be much easier for Abelian than for non-Abelian symmetry 
groups.)
For this reason, and for the sake of simplicity, we focus on
Simon- and Shor-type symmetries in the following.  

\vspace*{1pt}
\section{Simon-type symmetry}\label{Simon-type symmetry} 

As an expansion of Simon's problem with the periodicity condition
$f(x \oplus p)=f(x)$ we consider
\bea
\label{simon-symm}
V\left\{f(x),f(U[x])\right\}
=
f(x) \oplus f(x \oplus p) \oplus q = 0 
\;
\leadsto
\;
f(x \oplus p)
=
f(x) \oplus q
\,,
\ea
with $x,f(x),p,q\in\{0,1\}^n$, and the task is to find out $p$ and $q$.
For convenience, we shall identify bit-strings with integers
$\{0,1\}^n\leftrightarrow\{0,\dots,2^n-1\}$ via the usual binary 
representation in the following.
I.e., $x,f(x),p,q$ are treated as integers with $0 \leq x,f(x),p,q < N=2^n$.

In complete analogy to Simon's algorithm we apply the usual trick of inquiring
all entries at once (quantum parallelism) and obtain the state
\bea
\label{simon-run}
\ket{\Psi}=
\sum\limits_{\{x_0\}}^{(N/2)}
\frac{\ket{x_0}\ket{f(x_0)}+\ket{x_0 \oplus p}\ket{f(x_0) \oplus q}}{\sqrt{N}}
\,.
\ea
But instead of measuring the second register $\ket{f}$ we now perform a 
multiple application of the Hadamard gate to both, the first $\ket{x}$ 
and the second $\ket{f}$ register
\bea
\label{simon-trafo}
{\cal H}^{(2n)}\ket{\Psi}=
\frac{2}{\sqrt{N^3}}
\sum\limits_{\{x_0\}}^{(N/2)}
\sum\limits_{\{Y\,:\,R \cdot Y=0\}}^{(N^2/2)}
(-1)^{X \cdot Y}\ket{Y}
\,,
\ea
where we have introduced the abbreviations 
$\ket{X}=\ket{x_0}\otimes\ket{f(x_0)}$ and $\ket{R}=\ket{p}\otimes\ket{q}$ 
as well as the scalar product modulo two given by
\bea
\label{simon-result}
R \cdot Y = \sum\limits_{l=0}^{2n}R_lY_l\,{\rm mod}\,2 =
\bigoplus\limits_{l=0}^{n}\left(p_lY_l \oplus q_lY_{n+l}\right)
\,.
\ea
Assuming that the values $f(x_0)$ are pseudo-randomly distributed, i.e., 
without any internal order (cf.~the next Section), the measurement of
$Y$ returns arbitrary values satisfying the constraint $R \cdot Y=0$.
Again in complete analogy to Simon's algorithm, after $\ord(n)$ runs we have
enough measured values of $Y$ for determining $R$, i.e., $p$ and $q$, with 
arbitrarily high probability (exponential speed-up).

\vspace*{1pt}
\section{Requirements}\label{Requirements} 

In which cases can the above quantum algorithm fail, i.e., what
exactly does the aforementioned condition "without any internal order"  
imply? 

As a counter-example -- where the algorithm must fail -- consider the
function  
\bea
f(x)=\ul{A} \cdot x \oplus b
\,,
\ea
with a binary $N \times N$-matrix $\ul{A}$ and the bit-wise scalar
product modulo two as in  Eq.~(\ref{simon-result}).
This function exhibits a strong internal order and hence a plethora of
symmetries: any $p$ and the corresponding $q$ given by  
\bea
q=\ul{A} \cdot p
\,,
\ea
satisfies  Eq.~(\ref{simon-symm}).

On the other hand, as an example where the above quantum algorithm
works, we might construct the function $f(x)$ as follows:  
After splitting up the set of all arguments $\{x\}=\{0 \dots N\}$ into
two disjoint sets of equal strength $N/2$ via $\{x_0\}$ and 
$\{x_0 \oplus p\}$, we assign all $f(x_0)$ random values between 0 and
$N$ and determine the remaining ones via $f(x \oplus p)=f(x)\oplus q$.
In this rather artificial way we can make sure that there is no
additional internal order which could spoil the above algorithm.

In summary, we do not allow additional (exact or average) 
symmetries apart from the one in Eq.~(\ref{simon-symm}) which lead to 
another value $R' \neq R$ with the probability of measuring $R' \cdot Y=1$ 
being strongly suppressed.

Let us discuss the relation of the hidden symmetry discussed above to
the hidden sub-group problem. 
Defining new functions such as \cite{kitaev}
\bea
\label{kitaev}
h_1(x,y) = f(x) \oplus y
\,,
\;
h_2(x,y) = f(x) \oplus f(y)
\,,
\ea
the symmetry $f(x \oplus p)=f(x)\oplus q$ translates into periodicity  
\bea
h_1(x,y) = h_1(x \oplus p,y \oplus q)
\,,
\;
h_2(x,y) = h_2(x \oplus p,y \oplus p)
\,.
\ea
However, this identification does not imply that the property in
Eq.~(\ref{simon-symm}) can be mapped onto the hidden sub-group problem
as in Eq.~(\ref{subgroup}) because the functions 
$h_{1,2}\,:\,\{1,\dots,N^2\}\to\{1,\dots,N\}$ 
are highly degenerate and hence not distinct on different co-sets.

The fact that one can nevertheless find $p$ (and $q$) by a quantum algorithm 
(which is not necessary for such a large degeneracy) is caused by the
special underlying symmetry $f(x \oplus p)=f(x) \oplus q$ and the assumption
discussed above (no additional internal order).
Therefore, this is a true expansion of the hidden sub-group problem
\cite{kitaev} with the distinctness on different co-sets being
replaced by the pseudo-randomness requirement.

\vspace*{1pt}
\section{Shor-type symmetry}\label{Shor-type symmetry} 

As a second example for a hidden (two-point) symmetry, we study the
following expansion of Shor's problem $f(x+p)=f(x)$
\bea
\label{shor-symm}
f(x+p)=f(x)+q
\,,
\ea
with $0 \leq x,f(x) < N=2^n$.
Similar to the original period-finding algorithm, we demand that $p$
is much smaller than $N$, say $p=\ord(N^\varepsilon)$ with a small but 
positive number $0<\varepsilon<1$, which will be determined below.
In addition, we assume $p \gg q$ (but still $q \gg 1$) -- otherwise we
would have to insert a ``modulo $N$'', i.e., $f(x+p)=f(x)+q\mod N$.

In this situation, the usual superposition state after the application 
of the unitary gate calculating the function $f$ reads 
\bea
\label{shor-run}
\ket{\Psi}\approx
\sum\limits_{x_0=0}^{p-1}
\sum\limits_{l=0}^{[N/p]}
\frac{\ket{x_0+lp}\ket{f(x_0)+lq}}{\sqrt{N}}
\,,
\ea
where $[N/p]$ denotes the integer part of $N/p\gg1$ and the $\approx$
sign is caused by the corresponding neglect of a small number of
arguments $x$ and the fact that not all periods are complete 
(remember $p \gg q$). 

Again we do not measure the second register at this stage but apply a  
double quantum Fourier transform, i.e., we Fourier transform each
register 
\bea
\label{shor-trafo}
{\cal F}^{(2)}\ket{\Psi}
\approx
\sum\limits_{k_x=0}^{N-1}
\sum\limits_{k_y=0}^{N-1}
\sum\limits_{x_0=0}^{p-1}
\frac{e^{2\pi i(x_0k_x+f(x_0)k_y)/N}}{\sqrt{N^3}}
\sum\limits_{l=0}^{[N/p]}
\exp\left\{2\pi i\,\frac{pk_x+qk_y}{N}\,l\right\}
\ket{k_x}\ket{k_y}
\,.
\ea
Although the measurements of $k_x$ and $k_y$ considered separately
typically return almost random numbers -- provided that there is no 
structure (e.g., an additional periodicity, cf.~the previous example
as well as Sec.~\ref{Discrete self-similarity}) in the values $f(x_0)$ 
-- these numbers $k_x$ and $k_y$ display an extremely strong correlation: 
the above $l$-sum exhibits a constructive interference if and only if 
\bea
\label{shor-result}
\frac{pk_x+qk_y}{N}\in{\mathbb N}\pm\ord\left(\frac{p}{N}\right)
\ea
holds; and, accordingly, a large fraction of the measured values for 
$k_x$ and $k_y$ will obey this relation.

However (in contrast to Shor's algorithm) one measurement of $k_x$ and  
$k_y$ may not suffice for determining $p$ and $q$ in general.
To this end, it might be necessary to repeat the whole process a few times 
-- resulting in pairs of measured values $(k_x^a,k_y^a)$ with $a$ labelling 
the number of the measurement.
One possibility to derive $p$ and $q$ is to find a set of 
$A\in{\rm poly}(n)$ integers $\alpha_a\in{\mathbb Z}$ with 
$|\alpha_a| < M \ll N$ which satisfy
\bea
\label{cancel}
\sum\limits_{a=1}^A \alpha_a\,k_y^a \mod N = \ord(M)
\,.
\ea
Inserting the above condition back into Eq.~(\ref{shor-result}), we obtain 
(remember $p \gg q$)
\bea
\label{eliminate}
\frac{p}{N}\sum\limits_{a=1}^A \alpha_a\,k_x^a 
\in{\mathbb N}\pm\ord\left(\frac{ApM}{N}\right)
\,.
\ea
Having eliminated $q$ in this way, we may find $p$ via the continued fraction
expansion \cite{nielsen} of 
\bea
\label{fraction}
\xi=\frac{1}{N}\sum\limits_{a=1}^A \alpha_a\,k_x^a 
\in\frac{\mathbb N}{p}\pm\ord\left(\frac{AM}{N}\right)
\,,
\ea
provided that the denominator $p$ is small enough, i.e., 
$p \ll \sqrt{N}/\sqrt{AM}$.
There are two limits on the size of the auxiliary number $M$:
firstly, it should be small enough to allow the detection of sufficiently
large values of $p$ with $p \ll \sqrt{N}/\sqrt{AM}$, and, secondly, 
$M$ must be adequately large such that a small number of measured 
pairs $(k_x^a,k_y^a)$ will allow us to satisfy Eq.~(\ref{cancel}) 
with the probability that all of these pairs obey the resonance 
condition (\ref{shor-result}) not being exponentially suppressed.

For example, choosing $M=\sqrt{N}$, we may find $A=2$ numbers 
$|\alpha_1|<\sqrt{N}$ and $|\alpha_2|<\sqrt{N}$ via the continued 
fraction expansion of the ratio $k_y^1/k_y^2$ truncated at order 
$\sqrt{N}$ which then satisfy $\alpha_2/\alpha_1+k_y^1/k_y^2=\ord(1/N)$
and thus $\alpha_1k_y^1+\alpha_2k_y^2=\ord(\sqrt{N})$.
This allows us to find periods $p$ satisfying $p\ll\sqrt[4]{N}$ 
in two runs of the quantum algorithm with high probability. 
Note the difference of the above method to Shor's algorithm which 
requires $p\ll\sqrt{N}$ instead.

More generally, if $p=\ord(N^\varepsilon)$ is small enough 
(e.g., $\varepsilon<1/4$, see the above example), we are able 
to determine $p$ (i.e., $U$) and thereby also $q$ (i.e., $V$) 
in polynomial time (exponential speed-up).

\vspace*{1pt}
\section{Discrete self-similarity}\label{Discrete self-similarity}

Let us give an example where the above algorithm could be useful.
Starting from the Shor-type symmetry $f(x+p)=f(x)+q$ in Eq.~(\ref{shor-symm})
and setting 
\bea
f=\log(\phi)
\,,
\quad
x=\log(\chi)
\,,
\ea
with respect to some base(s), we arrive at
\bea
\label{self}
\phi(\alpha\,\chi)=\beta\,\phi(\chi)
\,,
\ea
i.e., the function $\phi(\chi)$ is discretely self-similar.
Discrete self-similarity -- also called discrete scale invariance -- 
is a characteristic feature of some non-linear systems 
(e.g., in condensed matter) exhibiting critical phenomena, 
see, e.g., \cite{self}.

For example, let us assume that the unitary gate ${\cal U}_\phi$ represents 
some characteristic parameter in the quantum simulation of a condensed matter 
system in the critical r\'egime and that this parameter $\phi$ displays a
discretely self-similar but otherwise chaotic dependence on some input 
$\chi$.
For the sake of simplicity, let us further assume that we can calculate 
the logarithms of the output $\phi$ and the input $\chi$ with respect to 
suitable bases within an appropriate discretisation
(either artificial or natural, e.g., physical lattice).
In this way the accomplished generalisation of pure periodicity 
$f(x+p)=f(x)\leftrightarrow\phi(\alpha\,\chi)=\phi(\chi)$ 
to discrete self-similarity in Eq.~(\ref{self}) in the presented quantum 
algorithm allows us to detect the discretely self-similar behaviour 
exponentially faster than any (known) classical method.

\vspace*{1pt}
\section{Summary}\label{Summary}

By means of a simple example, it has been demonstrated that the task of 
finding hidden (two-point) symmetries of a given function described by 
Eq.~(\ref{hidden}) -- as an expansion of the hidden subgroup problem in 
Eq.~(\ref{subgroup}) -- can also be accomplished exponentially faster 
by a (probabilistic) quantum algorithm than by classical methods.

There are two main possibilities for generating {\em NP}-problems
(i.e., the solution is potentially hard to find but  easy to verify, 
at least probabilistically) in this way -- either both, 
$U \leftrightarrow p$ and $V \leftrightarrow q$, are unknown or 
$V \leftrightarrow q$ is given and we have to find ``only'' 
$U \leftrightarrow p$ \cite{translation}. 
(Of course, if $p$ was known, the problem would be trivial.)

Note that the task under consideration is very similar to an inverse problem
where the input(s) and the output(s) of a function depending on a parameter 
are given and one has to find the fitting parameter.
We consider the main importance of our result in its being a small 
step towards the goal of better understanding the class of problems which can 
be solved exponentially faster by quantum algorithms.

\vspace*{1pt}
\section{Outlook}\label{Outlook} 

Eq.~(\ref{hidden}) does not represent the most general (explicit) two-point 
symmetry, which can be written as
\bea 
V\left\{x,f(x),f\left(U[x,f(x)]\right)\right\}=0
\,.
\ea
In this case there is no $f$-independent co-set in general
and it would be interesting to study the possibilities of speeding up these
more complicated (consistency, etc.) problems by quantum algorithms.
As another extension of Eq.~(\ref{hidden}), it appears quite natural to 
ask about relations involving more, say $(m+1)$, points 
\bea
\label{points}
V\left\{f(x),f(U_1[x]),\dots,f(U_m[x])\right\}=0
\,.
\ea

Further interesting symmetries\footnote{
As one possibility one might consider the case where the operations on the
argument (``inside'') and on the value of the function (``outside'') differ.
However, one must be careful:
For instance, if one defines the problem as, say, $f(x \oplus p)=f(x)+q$ or 
$f(x+p)=f(x) \oplus q$, the first case is inconsistent in general since
another iteration leads to a contradiction $f(x)=f(x)+2q$; and the second
example can be reduced to Shor's case  $f(x+2p)=f(x)$.} 
could include other transformations $U$ and relations $V$ -- think of
gauge symmetries, for example,  
or permutations (and other possibly non-Abelian groups).

Another point is that, in the examples considered above (and in the
hidden subgroup problem, of course), $V$ was invertible, i.e., one
could solve the relation  Eq.~(\ref{hidden}) for $f(x)$.
Relaxing this invertability condition would be another interesting
object of study.
As a very simple example, one might consider the following symmetry
\bea
\bigoplus\limits_{l=0}^{n} f_l(x) \oplus f_l(x \oplus p) = 0
\,,
\ea
where one can determine $p$ (again assuming appropriate conditions) 
via defining a new function $F(x)=\bigoplus\limits_{l=0}^{n} f_l(x)$.

\vspace*{1pt}
\section*{Acknowledgements}

The authors acknowledge valuable conversations with 
R.~Cleve, P.~H{\o}yer, A.~Kitaev and R.~Laflamme.
This work was supported by 
the Alexander von Humboldt foundation, 
the Canadian Institute for Advanced Research, 
the Natural Science and Engineering Research Council of Canada, and
the Pacific Institute of Theoretical Physics.
R.~S.~gratefully acknowledges financial support by the Emmy-Noether
Programme of the German Research Foundation (DFG) under grant 
No.~SCHU 1557/1-1,2.  

\vspace*{1pt}


\begin{thebibliography}{000}
\bibitem{intuitive}
As it becomes evident from the character of the arguments, they are still 
intuitive and suggestive and not mathematically rigorous or conclusive.
(Counter-examples are welcome.)
\end{thebibliography}
\end{document}